\documentclass[lettersize,journal]{IEEEtran}
\usepackage{amsmath,amsfonts}
\usepackage{algorithmic}
\usepackage{array}
\usepackage[caption=false,font=normalsize,labelfont=sf,textfont=sf]{subfig}
\usepackage{textcomp}
\usepackage{stfloats}
\usepackage{url}
\usepackage{verbatim}
\usepackage{graphicx}
\usepackage{hyperref}
\usepackage{amsmath}
\usepackage{amssymb}
\usepackage{mathtools}
\usepackage{amsthm}
\usepackage{stfloats}
\usepackage{cuted}
\usepackage{xcolor}
\usepackage{graphicx}    
\usepackage{subcaption}  
\usepackage{xcolor}      
\usepackage[capitalize,noabbrev]{cleveref}
\theoremstyle{plain}

\theoremstyle{definition}

\theoremstyle{remark}

\usepackage[textsize=tiny]{todonotes}
\usepackage{xcolor}         
\usepackage{amsmath}
\usepackage{color, colortbl}
\usepackage{graphicx}
\usepackage[most]{tcolorbox}
\usepackage{caption}
\usepackage{tikz}
\usepackage{subcaption}

\usepackage{wrapfig}
\usepackage[T1]{fontenc}    
\usepackage{url}            
\usepackage{booktabs}       
\usepackage{amsfonts}       
\usepackage{nicefrac}       
\usepackage{microtype}      
\usepackage{amsfonts,amssymb}
\usepackage{algorithm}
\usepackage{dsfont}
\usepackage{algorithmic}
\usepackage{multirow}
\usepackage{pbox}
\usepackage{bbm}
\definecolor{LightCyan}{rgb}{0.88,1,1}

\newcommand{\nsep}{{\textbackslash}n{\textbackslash}n}

\usepackage[most]{tcolorbox}

\def\BibTeX{{\rm B\kern-.05em{\sc i\kern-.025em b}\kern-.08em
    T\kern-.1667em\lower.7ex\hbox{E}\kern-.125emX}}
\usepackage{balance}
\begin{document}

\title{Holistic Utility Preference Learning for Listwise Alignment}

\author{Jiacong Zhou, Xianyun Wang, Min Zhang, Jun Yu
\thanks{Jiacong Zhou is with the School of Computer Science and Technology, Hangzhou Dianzi University, Hangzhou 310018, China (e-mail: jczhou@hdu.edu.cn).}

\thanks{Xianyun Wang, Min Zhang, and Jun Yu are with the Harbin Institute of Technology, Shenzhen 518055, China (e-mail: wangxianyun627@gmail.com; minzhang@suda.edu.cn; yujun@hit.edu.cn).}
}


\maketitle

\begin{abstract}
Aligning large language models with human preferences is essential for improving interaction quality and safety by ensuring outputs better reflect human values. A promising strategy involves Reinforcement Learning from Human Feedback (RLHF), starting with collecting and ranking responses generated by a supervised fine-tuning model to refine alignment. Existing methods such as Direct Preference Optimization (DPO) focus on pairwise comparisons, categorizing responses into preferred and less preferred pairs and optimizing pairwise margins. However, this pairwise approach cannot capture the holistic ranking relationships among multiple responses or effectively leverage the rich preference information available in list-wise comparisons. To address this challenge, this paper introduces \underline{D}irect \underline{R}anking \underline{P}reference \underline{O}ptimization (DRPO), a novel method that views human preference alignment as a Learning-to-Rank (LTR) task. Unlike pairwise methods, DRPO optimizes the preference ranking of entire response lists by computing holistic utility scores through NDCG, a standard LTR metric. To enable end-to-end optimization with the non-differentiable NDCG, we propose diffNDCG loss, a differentiable approximation facilitated by a sorting network. Furthermore, we introduce a novel margin-based Adaptive Rank Policy Score to enhance the discriminative quality of generated responses. Extensive experiments have shown that DRPO outperforms existing methods, enhancing the quality of the generated responses. The code is available \href{https://github.com/choucaicai/drpo-align}{here}
\end{abstract}

\begin{IEEEkeywords}
Preference Alignment, Learning-to-Rank, Large Language Models
\end{IEEEkeywords}

\section{Introduction}
\IEEEPARstart{L}{arge}
language models (LLMs), trained on extensive and diverse datasets, can be prompted to demonstrate impressive capabilities across a broad range of tasks \cite{huang2024good,vicuna2023,openai2024gpt4,touvron2023llama,DBLP:journals/taslp/ZhuQFCHX24,DBLP:journals/taslp/NiLYK24}. However, due to the varied nature of their training data, these models sometimes produce content that may not align with human preferences, including fabricated answers, offensive comments, or harmful responses \cite{bai2022training,DBLP:conf/nips/WangWSL23}. To ensure the development of AI systems that are safe and controllable, this paper investigates learning tasks for LLMs that guide them to generate responses in alignment with human preferences. 

Human preference alignment has emerged as an active research area, pioneered by Reinforcement Learning from Human Feedback (RLHF) \cite{ouyang2022training}. However, the RLHF optimization process is complex and its implementation introduces challenges due to unstable and costly training. Recent studies \cite{hong2024orpo,ethayarajh2024kto} have started to adopt alternatives to RLHF. For example, Direct Preference Optimization (DPO) \cite{rafailov2023direct} enables the extraction of the corresponding optimal policy in a closed form and derives a pairwise logistic loss directly from pairwise preference data. DPO eliminates the need for explicit reward modeling or reinforcement learning, thereby reducing the training costs associated with RLHF. 


Although significant progress has been made in human preference alignment, most existing methods primarily focus on \textit{pairwise human preferences}, which involve evaluating human preferences by comparing \textit{preferred} and \textit{less-preferred} responses. Nevertheless, human preferences are not solely expressed as preferences and less preferences; they also manifest as ranking information, an aspect that has rarely been explored in previous research. In practice, \textit{ranking preference data} is  widely utilized. For instance, in the Ultrafeedback and VLfeedback datasets \cite{2023vlfeedback,cui2023ultrafeedback}, multiple responses are generated using a Supervised Fine-tuning (SFT) model. These responses are then evaluated and ranked by leveraging advanced AI technologies such as GPT-4 \cite{openai2024gpt4}.
Moreover, as demonstrated by \cite{liu2024lipo}, presenting ranking preference data effectively distribute the costs associated with processing the prompt \cite{liu2023statistical}. 

In this work, we approach human preference alignment as a listwise ranking problem. While straightforward solutions extend existing pairwise alignment methods to ranking scenarios, they fail to fully utilize the relative preference strengths within ranked lists \cite{DBLP:conf/acl/ZhuLZGM24}. Recent studies have attempted to address this limitation through two main strategies: utilizing the Plackett-Luce preference model or employing pairwise methods with listwise-aware weighting schemes \cite{yuan2023rrhf, chen2024preference, DBLP:conf/acl/ZhuLZGM24, DBLP:conf/icml/ChoiJAM24, liu2024lipo}.

However, a \textit{mismatch} persists between evaluation metrics and optimization objectives in all current methods. Alignment quality is typically assessed using \textit{win rates}, which measure how often one model's responses are preferred over another's, yet existing alignment methods do not directly optimize this criterion. This mismatch means that optimizing current loss functions \cite{liu2024lipo,song2023preference} may not necessarily lead to higher win rates or improved human preference satisfaction \cite{chen2024preference}. While traditional learning-to-rank tasks have demonstrated that directly optimizing evaluation metrics is highly effective \cite{Pobrotyn2021NeuralNDCGDO,DBLP:journals/ir/QinLL10}, and recent studies \cite{chen2024preference} reveal correlations between win rates and ranking accuracy, current alignment methods achieve low ranking accuracy on standard datasets. The relationship between win rates and more comprehensive ranking metrics like NDCG remains unexplored, making direct optimization of evaluation criteria a critical yet challenging research gap.

To tackle this challenge, we propose \underline{D}irect \underline{R}anking \underline{P}reference \underline{O}ptimization (DRPO), a novel method that directly aligns LLMs with Normalized Discounted Cumulative Gain (NDCG), a standard Learning-to-Rank metric that accurately quantifies ranking quality. Specifically, we introduce the \textit{Adaptive Rank Policy Score}, a ranking strategy that maximizes the likelihood of preferred responses while dynamically adjusting score margins based on relative positions in the ranked list.
Furthermore, we implement \textit{differentiable sorting networks} to sort responses based on computed scores. This sorting method combines implementation simplicity and computational efficiency, while yielding doubly stochastic permutation matrices that preserve probability distributions and enable efficient differentiable optimization \cite{petersen2021differentiable}. Additionally, due to the nondifferentiability of NDCG, we leverage permutation matrices develop a differentiable version of NDCG (i.e., \textit{diffNDCG}) to serve as the loss function, simulating the NDCG metric. Optimizing the diffNDCG loss enhances model performance by prioritizing top-ranked responses and imposing stricter penalties for misplacing highly relevant items, without additional computational overhead compared to existing methods.
The main contributions of our work are summarized as follows:

$\bullet$ A novel method Direct Ranking Preference Optimization (DRPO) is developed to explore listwise human preference alignment with ranking preference data.

$\bullet$ A novel ranking score computation strategy, the Adaptive Rank Policy Score, has been introduced, which maximizes the likelihood of preferred responses while dynamically adjusting score margins based on relative positions.

$\bullet$  A novel differentiable NDCG (diffNDCG) metric has been developed to emulate the NDCG metric used in LTR. By optimizing diffNDCG, we prioritize top-ranked responses while penalizing their misplacement in lower positions.

\section{Related Works}

\textbf{Human Preference Alignment.}
Although LLMs demonstrate impressive capabilities, they sometimes produce harmful content, leading to safety concerns. To address this, Reinforcement Learning with Human Feedback (RLHF) \cite{bai2022training, DBLP:conf/nips/ChristianoLBMLA17} aligns LLMs with human preferences by training a reward model using the Bradley-Terry model \cite{david1963method} and fine-tuning with reinforcement learning algorithms like PPO \cite{schulman2017proximal}. However, RLHF suffers from instability and high computational costs. To reduce reliance on reinforcement learning, Direct Preference Optimization (DPO) \cite{rafailov2023direct} derives a closed-form optimal policy with pairwise logistic loss, eliminating explicit reward modeling. RSO \cite{liu2023statistical} employs rejection sampling for more accurate policy estimation and provides a unified framework extending SLiC \cite{zhao2023slic} and DPO.

Most recent works  \cite{cheng2023adversarial,DBLP:conf/aistats/AzarGPMRVC24,yuan2023rrhf,DBLP:journals/taslp/KananAAHMKK24} focus on pairwise preferences, with limited attention to listwise settings. RRHF \cite{yuan2023rrhf} uses pairwise hinge loss but treats pairs independently rather than modeling the complete ranking structure. PRO \cite{song2023preference} employs the Plackett-Luce model \cite{af5079a1-8ca5-3727-a405-0a82390327b7} with list MLE loss. LiPO \cite{liu2024lipo} frames preference alignment as learning-to-rank and uses DCG-weighted pairwise logistic loss (lambda loss) \cite{47258} to indirectly optimize NDCG. However, \cite{Pobrotyn2021NeuralNDCGDO} showed that direct NDCG optimization yields superior performance in traditional learning-to-rank tasks.

Building upon this insight, we implement differentiable sorting networks to obtain differentiable permutation matrices of responses. It enable us to develop a differentiable NDCG(diffNDCG), allowing direct optimization of NDCG for ranked lists to more effectively align LLMs with human preferences.

\textbf{Learning-to-Rank Task.} Learning-to-Rank (LTR) task is a well-studied field with extensive literature, primarily due to its practical applications in web search and recommendation systems \cite{liu2009learning,DBLP:conf/sigir/DaiSD11,macdonald2013whens,cao2007learning,yu2013exploiting,yu2019hierarchical,abdollahpouri2017controlling,DBLP:conf/aaai/ZhangFZYT0TDY24}. Traditional research in Learning to Rank (LTR) has concentrated on developing robust ranking objectives, including pointwise, pairwise, and listwise approaches. RankSVM \cite{cao2006adapting} and RankNet \cite{burges2005learning} utilize pairwise hinge loss and pairwise logistic loss, respectively, to optimize ranking performance. ListMLE and Softmax losses are two representative listwise losses introduced by \cite{cao2007learning}. LambdaRank \cite{burges2006learning} employs a pairwise logistic loss with lambda weights and achieves strong empirical performance improvements compared to RankNet. Furthermore, numerous methods have been proposed to directly optimize the non-smooth NDCG metric. For instance, SoftRank\cite{Taylor2008SoftRankON},ApproxNDCG~\cite{DBLP:journals/ir/QinLL10} employs rank distributions to smooth NDCG. PiRank \cite{DBLP:conf/nips/SwezeyGCE21} and NeuralNDCG \cite{ Pobrotyn2021NeuralNDCGDO} approximate the non-continuous sorting operator using NeuralSort \cite{DBLP:conf/iclr/GroverWZE19} to smooth NDCG. Inspired by these studies, we utilize differentiable sorting networks to smooth NDCG and introduce the diffNDCG loss function for preference alignment.

\section{Preliminaries}

\textbf{Prompt, Response and Policy.} Let $\mathcal{X}$ and $\mathcal{Y}$ denote the set of prompts and the set of responses (action space), respectively. We use $x \in \mathcal{X}$ to represent a prompt, and $y \in \mathcal{Y}$ to represent a response. Given a prompt $x$, a large language model (LLM) generates a corresponding response $y$. This response $y$ is produced according to a policy $\pi_{\theta}(\cdot|x)$, which is a discrete distribution over $\mathcal{Y}$. We also define $\pi_{\text{ref}}(\cdot|x)$ as a discrete distribution over $\mathcal{Y}$, serving as the reference policy. The reference policy $\pi_{\rm ref}$ is derived from the Supervised Fine-tuning (SFT) model \cite{rafailov2023direct}.

\textbf{Ranking Preference Data.} The training dataset $D = \{x^i, \mathbf{y}^i, \mathbf{s}^i\}_{i=1}^N$ is composed of three elements: $x^i$ represents the $i$-th prompt; $\mathbf{y}^i = (y^i_1, \ldots, y^i_K)$ consists of a list of $K$ responses, typically generated by the SFT model; and $\mathbf{s}^i = (s^i_1, \ldots, s^i_K) \in [0,1]^K$ denotes the relevance scores of the responses $\mathbf{y}^i$ in relation to the prompt $x^i$. The relevance score $s_j^i$, generally obtained from AI \cite{huang2024good,jiang2023mistral,bai2023qwen,vicuna2023} and human feedback or a reward model, reflects how well the response $y_j^i$ corresponds to the prompt $x^i$. If the response $y^i_j$ is scored higher than $y^i_l$, it implies that $y^i_j$ is more closely aligned with human preferences compared to $y^i_l$.


\textbf{Alignment with Human Preferences Using Ranking Preference Data.} Aligning LLM with human preferences involves utilizing the dataset of human preferences (e.g., the training dataset $D$) to refine the policy $\pi_{\theta}(y|x)$. Substantial progress has been made toward achieving this goal, with most existing studies \cite{rafailov2023direct,hong2024orpo} leveraging \textit{pairwise} preference data, represented as $D_P = \{x^i, (y^i_1,y^i_2), (s^i_1,s^i_2)\}_{i=1}^N$ (i.e., the case $K=2$ for training dataset $D$). Unlike existing methods, we treat preference alignment as a \textit{listwise ranking} problem (i.e., $K\geq2$), which allows for a more effective exploitation of the complex preference relationships embedded within sequence.

\textbf{Learning-to-Rank Task.}
When a user enters a query $x_q$, the LTR algorithm \cite{liu2009learning,yu2019hierarchical,cao2007learning} aims to accurately rank candidate documents (including texts, images, and other modalities) by relevance, prioritizing the most pertinent information for efficient user retrieval. Let $\mathbf{M}$ denote the LTR model that assigns relevance scores $\mathbf{s}$ between a query ${x}_q$ and documents $\mathbf{y}_d$, where higher scores indicate more relevant documents.

\textbf{Learning Framework.} Let $\mathbf{M}$ be the score prediction model  of human preference alignment. Given the prompt $x$ and responses $\mathbf{y}$, one can predict the relevance scores:
\begin{equation*}
    \hat{\mathbf{s}}_{\theta} = \mathbf{M}(x,\mathbf{y}; \pi_{\theta}),
\end{equation*}
where $\pi_{\theta}$ is the LLM policy and $\theta$ is the corresponding parameters. Let $\ell : (\mathbf{s},\hat{\mathbf{s}_{\theta}})  \rightarrow \mathbb{R}$ be the loss function. We will learn the parameters $\theta$ based on the empirical risk minimization principle:
\begin{equation}\label{all_loss}
    \theta \in \texttt{argmin}_{\theta} ~{\mathcal{L}}(\theta) = \frac1{|D|}\sum_{(x,\mathbf{y},\mathbf{s})\in D} \ell (\hat{\mathbf{s}}_{\theta},\mathbf{s}).
\end{equation}

\begin{figure*}
    \centering
    \includegraphics[width=\textwidth]{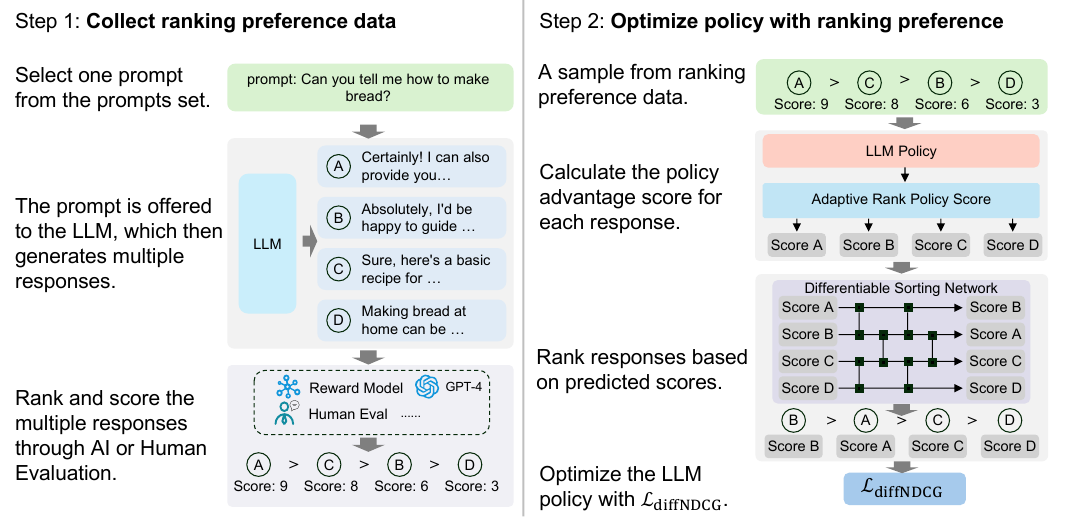}
    \caption{Illustration of two key steps of our method: (1) collecting ranking preference data, and (2) optimizing the policy with the collected ranking preference data. In Step 2, the Adaptive Rank Policy Score computes the score for each input response. Subsequently, the Differentiable Sorting Network sorts these responses based on their scores. We then compute the diffNDCG between the predicted scores and the ground-truth scores, and optimize the policy using the diffNDCG loss.}
    \label{fig:overview}
    \vspace{-1.25em}
\end{figure*}

\section{Proposed Methodology}
\label{section:three}
In this section, we introduce the proposed method DRPO, which comprises three primary components: (1) \textit{ranking score computation}; (2) \textit{differentiable responses ranking};  (3) \textit{diffNDCG loss}. Figure \ref{fig:overview} shows the overview of our method.

\subsection{Ranking Score Computation}
\label{sec:score_comp}r
\textbf{Policy Reference
Ratio.} The fundamental criterion for computing the ranking score is that more preferred responses should receive higher scores.  A commonly used strategy to compute the ranking scores is \textit{Policy Reference Ratio} 
proposed by \cite{rafailov2023direct}: Let $\mathbf{M_{prr}}$ denote the policy reference ratio model for human preference alignment, which is defined as:
\begin{equation}
    \begin{gathered}
        \mathbf{M_{prr}}(x,\mathbf{y}; \pi_{\theta}) =  \Big( \beta\log\frac{\pi_\theta(y_1 \mid x)}{\pi_{\text{ref}}(y_1 \mid x)},...,\\ \beta\log\frac{\pi_\theta(y_K \mid x)}{\pi_{\text{ref}}(y_K \mid x)} \Big),
    \end{gathered}
    \label{m1_dpo}
\end{equation}
where $\beta$ denotes the regularization coefficient controlling the KL divergence between $\pi_{\theta}$ and $\pi_{\text{ref}}$.

\textbf{Adaptive Rank Policy Score.} While the \textit{Policy Reference Ratio} defined in Eq. \ref{m1_dpo} has been widely adopted in various methods \cite{liu2024lipo}, it emphasizes the relative likelihood between the policy model $\pi_{\theta}$ and a reference model $\pi_{\mathrm{ref}}$ rather than directly maximizing the absolute likelihood of the preferred response. Consequently, a high Policy Reference Ratio score may coincide with low absolute likelihood for preferred responses \cite{meng2024simpo}, leading to sub-optimal performance in real-world generation tasks,  where high absolute likelihoods are essential for producing high quality outputs \cite{holtzman-etal-2018-learning,fan-etal-2018-hierarchical}. To address this, we focus on the log-likelihood of generated sequences and establish a length normalized basic scores function based on log-likelihood:
\begin{equation}
    \begin{gathered}
        \label{s_prob}
        \mathbf{s}(x,\mathbf{y} ; \pi_{\theta}) = \Big (
            \frac{1}{|y_1|} \log \pi_\theta(y_1 \mid x),...,\\ \frac{1}{|y_K|} \log \pi_\theta(y_K \mid x)
        \Big),
    \end{gathered}
\end{equation}
Here, $|y|$ denotes the token length of $y$. This length normalization reduces bias towards shorter sequences \cite{yuan2023rrhf}. Furthermore, in the process of ranking responses based on their scores, pairwise score differences between elements are computed. A common practice to enhance discrimination between high and low-quality responses is to incorporate a margin \cite{meng2024simpo,ethayarajh2024kto}, ensuring preferred responses exceed dispreferred ones by at least a specified threshold. Empirical evidence suggests that margin-based approach improves both model generalization and response quality \cite{touvron2023llama}. Therefore, we further introduce a ranking-aware term $\gamma(y)$.
\begin{equation}
    \begin{gathered}
        \label{s_prob_gamma}
        \mathbf{s}(x,\mathbf{y} ; \pi_{\theta}) = \Big (
                \frac{1}{|y_1|} \log \pi_\theta(y_1 \mid x) + \gamma(y_1),..., \\  \frac{1}{|y_K|} \log \pi_\theta(y_K \mid x) + \gamma(y_K)
             \Big ).
    \end{gathered}
\end{equation}
The ranking-aware margin is then defined as the difference between $\gamma(y_i)$ and $\gamma(y_j)$ when comparing scores of two responses $y_i$ and $y_j$. This margin should effectively reflect  quality differences among responses across the ranked list. Specifically, we assume adjacent responses have similar relevance, and design ranking-aware margin to satisfy three criteria: first, apply smaller margins for adjacent ranks, allowing fine-grained discrimination; second, assign larger margins for greater ranking disparities, emphasizing significant differences; finally, dynamically adjust margins based on relative score changes, maintaining discrimination across the quality levels while avoiding overemphasis on minor differences. Therefore, we define $\gamma(y)$ by combining a base weighted rank term with an exponential moving average estimate \cite{Qin2010AGA} of rank-related past scores:
\begin{equation}
\label{eq:gamma_y_2}
\begin{aligned}
\gamma(y) &= \tau \cdot q(y) - \beta \cdot {V}_{q(y)},~
\text{where} ~ \\ {V}_{q(y)} &\gets \theta \cdot {V}_{q(y)} + (1-\theta) \cdot \frac{1}{|y|} \log \pi_\theta(y \mid x).
\end{aligned}
\end{equation}
Here, $q(y)$ is the rank of response $y$ ($0$ for the top-ranked response), and $\tau$ is a positive constant denoting the basic margin between adjacent responses. ${V}_{q(y)}$ is the exponential moving average estimate of the log likelihood at rank $q(y)$ for dynamically tracking historical changes. $\theta \in [0, 1]$ is the parameter controlling the update rate, while $\beta$ determines the influence of the historical estimate on the current score. Based on the adaptive ranking-aware term $\gamma(y)$, we propose the novel \textit{ adaptive ranking policy score}:
\begin{equation}
    \label{M_ad}
    \begin{aligned}
        \mathbf{M}_{\rm arp}(x,\mathbf{y}; \pi_{\theta}) &=  \Big ( 
            s(x,y_1 ; \pi_{\theta}),...,s(x,y_K ; \pi_{\theta})
            \Big ), 
    \end{aligned}
\end{equation}
where $s(x,y ; \pi_{\theta})= \log {\pi_\theta(y \mid x)}/{|y|} + \tau \cdot q(y) - \beta V_{q(y)}$. Detailed and rigorous experiments demonstrate that our Adaptive Ranking Policy Score consistently outperforms the Policy Reference Ratio score across a wide range of metrics and diverse datasets (see Section \ref{sec:exp}).

\subsection{Differentiable Responses Ranking}

One of the most intuitive strategies to learn human preferences from the response list $\mathbf{y}$, is to sort the responses by predicted scores and use this ranking to fine-tune the language model, thereby learning the optimal preference ordering.
However, traditional sorting methods (e.g. Selection Sort \cite{musser1997introspective}, Quick Sort \cite{hoare1962quicksort}) operate in a discontinuous manner, impeding differentiable optimization in LLM fine-tuning for preference learning \cite{petersen2021differentiable}. In this section, we employ a \textit{differentiable sorting network}\footnote{Sorting networks are specialized computational architectures that sort sequences through comparisons and exchanges, not neural networks for sorting tasks.} to rank responses based on scores $\hat{\mathbf{s}}_{\theta}$, enabling end-to-end fine-tuning of LLM on ranking preferences.

Differentiable sorting networks offer superior parallel efficiency and excellent sorting performance while maintaining differentiability \cite{petersen2021differentiable}. For a list of length $L$, the time complexity ranges from $\mathcal{O}(L^2)$ to $\mathcal{O}(L \log^2{L})$, depending on the specific network variant \cite{10.5555/533017} (such as Odd-Even or Bitonic networks) \cite{10.5555/533017}. These complexities are competitive with or surpass many differentiable sorting methods, including those proposed in \cite{song2023preference,DBLP:conf/iclr/GroverWZE19,liu2024lipo,DBLP:conf/icml/XiaLWZL08,DBLP:conf/icml/BlondelTBD20,DBLP:conf/nips/SwezeyGCE21}. A comprehensive time complexity comparison is provided in Table \ref{tab:hh-qwen-cost}. Furthermore, sorting networks produce doubly stochastic permutation matrices, crucial for accurate NDCG computation by representing ranking probabilities faithfully. In contrast, existing differentiable sorting methods \cite{DBLP:conf/iclr/GroverWZE19,DBLP:conf/nips/SwezeyGCE21} often produce unimodal permutation matrices, leading to overestimation in response gain calculations. The experimental results in Table \ref{tab:hh-qwen} demonstrate that sorting networks significantly outperform existing differentiable sorting methods in various metrics.

\textbf{Odd-even Sorting Network.} In this work, we adopt the odd-even sorting network \cite{batcher1968sorting} for response ranking, which shows promising performance in our experiments while maintaining implementation simplicity and efficiency. Given the predicted scores $\hat{\mathbf{s}}_{\theta} = \mathbf{M}_{\rm arp}(x,\mathbf{y}; \pi_{\theta}) = (\hat{s}_1, \ldots, \hat{s}_K)$ (as defined in Eq. \ref{M_ad}), we employ a $K$-layer odd-even sorting network to sort these scores in descending order. This sorting network alternates between odd and even indices. In each layer, selected elements $\hat{s}_j$ are compared with their subsequent elements $\hat{s}_{j+1}$ and swapped if not in descending order. This process continues through $K$ layers until sequence $\hat{\mathbf{s}}_{\theta}$ is sorted. Appendix  presents a comprehensive procedure of this sorting network.


\textbf{Differentiable Swapping Operation.} Generally, For adjacent pairs $(j, j+1)$ where $j \in {1,...,K-1}$, the swapping operation is defined as:
\label{sec:diff_swap_op}
\begin{equation*}
\hat{s}'_j = \max(\hat{s}_j,\hat{s}_{j+1}),~~~\hat{s}'_{j+1} = \min(\hat{s}_j,\hat{s}_{j+1}),
\end{equation*}
Since the operations of $\max$ and $\min$ are non-differentiable, we need to modify the swapping operation to ensure the ranking process is differentiable. Following
\cite{petersen2021differentiable,petersen2022monotonic}, we can refine the $\min$ and $\max$ as follows:
\begin{equation}
\label{eq:diff_swap}
\begin{aligned}
        {\min}_{\rm soft}(\hat{s}_j,\hat{s}_{j+1}) &= \hat{s}_j \cdot  \alpha  + \hat{s}_{j+1} \cdot (1-\alpha), \\ {\max}_{\rm soft}(\hat{s}_j,\hat{s}_{j+1}) &= \hat{s}_j \cdot (1-\alpha)  + \hat{s}_{j+1} \cdot   \alpha, \\
        \alpha &= h(\hat{s}_{j+1}-\hat{s}_j), 
\end{aligned}
\end{equation}
where $h(\cdot)$ is a $s$-shaped function, which can be defined as:
\begin{equation}
\label{eq:stepness}
h(x)=
\begin{cases}
-\frac{1}{16\alpha x} \quad \quad  \text{if} ~\alpha x < -0.25,\\
1 - \frac{1}{16 \alpha x} \quad  \text{if} ~\alpha x > 0.25, \\
\alpha x + 0.5  \quad \text{otherwise},
\end{cases}
\end{equation}
here $\alpha$ represents the steepness that controls the relaxation strength. Then, we can reformulate the differentiable swapping operations described in Eq. \ref{eq:diff_swap} for the $k\text{-th}$ layer using a $K \times K$ permutation matrix, denoted as $\mathbf{P}_k$. Specifically, a swapping operation at index $j$ during either the odd or even stage can be represented as $[\hat{s}'_{j}, \hat{s}'_{j+1}] =  [\hat{s}_j, \hat{s}_{j+1}]\cdot \mathbf{p}_j$, where $\mathbf{p}_j$ is a $2 \times 2$ matrix defined as:
\begin{equation*}
    \mathbf{p}_j = 
    \begin{bmatrix}
        1-h(\hat{s}_{j+1}-\hat{s}_{j}) & h(\hat{s}_{j+1}-\hat{s}_{j})  \\
        h(\hat{s}_{j+1}-\hat{s}_{j}) & 1 -h(\hat{s}_{j+1}-\hat{s}_{j})\\
    \end{bmatrix}.
\end{equation*}
To integrate all swapping operations in the $k\text{-th}$ layer, we aggregate the matrices corresponding to all odd or even indices, yielding the permutation matrix of the $k$-th layer.
\begin{equation}
    \label{eq:P_soft}
    \mathbf{P}_k = {\rm diag}(\mathbf{p}_1, \mathbf{p}_3, \ldots) ~~\text{or}~~ {\rm diag}(1, \mathbf{p}_2, \mathbf{p}_4, \ldots).
\end{equation}
Multiplying the permutation matrices of all layers yields the overall permutation matrix $\mathbf{P}_{\text{soft}} = \mathbf{P}_1 \cdot \ldots \cdot \mathbf{P}_K$. The final ordered scores are given by $\hat{\mathbf{s}}_{\text{order}} = \mathbf{P}_{\text{soft}}^{\top} \cdot \hat{\mathbf{s}}_{\theta}$.

\subsection{Differentiable Normalized Discounted Cumulative Gain Loss}

To align human preferences with differentiable sorting, a direct strategy involves optimizing the cross-entropy loss  between the ground truth permutation matrix $\mathbf{P_{\rm ground}}$ (obtained by ground truth scores $\mathbf{s}$) and the predicted soft permutation matrix $\mathbf{P}_{\rm soft}$: let $ \ell_{\rm ce}$ be the cross-entropy loss and $[\mathbf{P}]_j$ be the $j$-th column of matrix $\mathbf{P}$,
\begin{equation}
    \label{eq:loss_ce}
    \mathcal{L}_{\rm ce} =\frac{1}{K} \sum_{j=1}^{K} \ell_{\rm ce}([\mathbf{P}_{\rm soft}]_{{j}},[\mathbf{P}_{\rm ground}]_{{j}}).
\end{equation}
However, experiments in Table \ref{tab:ablation} reveals this strategy's suboptimality. One possible explanation is that it fails to distinguish error severity across ranking positions, incorrectly equating misplacements of top-ranked responses with lower-ranked items, despite higher-ranked responses typically being far more crucial. To address these challenges, we propose optimizing Normalized Discounted Cumulative Gain (NDCG) \cite{Jrvelin2002CumulatedGE}, an effective LTR metric for measuring ranking quality. NDCG assesses the significance of responses in conjunction with their ranking positions.
It assigns greater importance to responses at the top of the ranking compared to those positioned
lower, and imposes a more severe penalty for inaccurately placing a top-ranked response in a lower
position.

Furthermore, while NDCG was originally designed to reflect users' tendency to focus on top-ranked results \cite{Jrvelin2002CumulatedGE}, this characteristic aligns with human preference rankings, which prioritize more preferred responses over less preferred ones \cite{pool2016attentional}. This similarity makes NDCG \textit{an effective proxy for evaluating and learning human preferences}. Experiments in Figure \ref{fig:ndcg-winrate} demonstrate that has a stronger correlation with human preference win rates compared to the optimizing targets of existing methods, highlighting its effectiveness. Additionally, NDCG prioritizes top-ranked responses and penalizes their misplacement, and \textit{capturing graded importance between responses without introducing any additional computation burden}. Table \ref{tab:hh-qwen-cost} shows our method achieves comparable computational complexity to existing methods.

\textbf{Normalized Discounted Cumulative Gain.} For a data point $(x, \mathbf{y}, \mathbf{s})$, NDCG can be written as:
\begin{equation}
    \label{eq:cal_ndcg}
\begin{aligned}
     \mathrm{NDCG}(\hat{\mathbf{s}}_{\theta},\mathbf{s}) &=\frac{1}{\rm iDCG}\sum_{j=1}^K\frac{2^{s_j}-1}{\log_2(1+q(y_j))},  \\  ~\text{where} ~{\rm iDCG} &= \sum_{j=1}^K \frac{{2^{s_j}-1}}{{\log_2(1+q^*(y_j))}},
\end{aligned}
\end{equation}
Here $q(y_j)$ and $q^*(y_j)$ denote the ranking positions of $y_j$ with respect to $\hat{\mathbf{s}}_{\theta}$ and $\mathbf{s}$, respectively.
NDCG assigns gains of $2^{s_j} - 1$ based on the score $s_j$ of each response $y_j$. It also applies discount factors ${\log_2(1 + q(y_j))}$ to each response according to its rank $q(y_j)$. This discount assigns higher weights to elements at the top of the ranking. Consequently, NDCG effectively accounts for the varying importance of responses at different ranking positions. However, the required sorting operation in NDCG makes it non-differentiable with respect to ranking positions $q(y_j)$.

\begin{algorithm}[t]
    \caption{Direct Ranking Preference Optimization (DRPO).}
    \label{algorithm:drpo}
    
    \renewcommand{\algorithmicrequire}{\textbf{Input:}}
    \renewcommand{\algorithmicensure}{\textbf{Output:}}
    \renewcommand{\algorithmiccomment}[1]{\hfill $\triangleright$ #1}

    \begin{algorithmic}[1]
        \STATE \textbf{Data:} Ranking preference data $ D = \{x^i, \mathbf{y}^i, \mathbf{s}^i\}_{i=1}^N $.
        \STATE \textbf{Initialize:} Policy $\pi_{\theta}$, reference policy $\pi_{\rm ref}$ .
        
        \FOR{ sample a batch $ B = \{x, \mathbf{y}, \mathbf{s}\} \subset D$ }
            \STATE Obtain predicted score \(\hat{\mathbf{s}}_{\theta} = \mathbf{M}_{\rm ad}(x, \mathbf{y}, \pi_{\theta})\) for each data point form $B$.
            \STATE Perform differentiable sorting based on score \(\hat{\mathbf{s}}_{\theta}\) and obtain the differentiable permutation matrix \(\mathbf{P}_{\rm soft}\).
            \STATE Calculate the substitute score \(\psi'(d, \mathbf{s}, \hat{\mathbf{s}}_{\theta}) = [\mathbf{P}_{\rm soft}^\top \cdot \mathbf{s}]_d\) for each data point form $B$.
            \STATE Calculate the diffNDCG for each data point form $B$:
            \[
                 \mathrm{diffNDCG}(\hat{\mathbf{s}}_{\theta},\mathbf{s}) =\frac{1}{\rm iDCG}\sum_{d=1}^K\frac{2^{\psi '(d,\mathbf{s},\hat{\mathbf{s}}_{\theta})}-1}{\log_2(1+d)}.
            \]
            \STATE Use gradient descent to update the parameters \({\theta}\) of following objective function:
            $$
             \mathcal{L}_{\rm diffNDCG} = \frac1{|B|} \sum_{(x,\mathbf{y},\mathbf{s})\in B} \ell_{\rm diffNDCG}(\mathbf{M}_{\rm ad}(x,\mathbf{y},\pi_{\theta}),\mathbf{s}).
            $$

        \ENDFOR
        
        \STATE \textbf{Return:} Policy \(\pi_{\theta}\)
    \end{algorithmic} 
\end{algorithm}

\textbf{Differentiable Normalized Discounted Cumulative Gain.} In this work, we introduce the \textit{Differentiable Normalized Discounted Cumulative Gain} (diffNDCG), which reformulates NDCG using a differentiable sorting mechanism.
We first reformulate Eq. \ref{eq:cal_ndcg}: let $d=q(y_j)$ and $\psi(d,\mathbf{s},\hat{\mathbf{s}}_{\theta})=s_j$,
\begin{equation}
    \begin{aligned}
         \mathrm{NDCG}(\hat{\mathbf{s}}_{\theta},\mathbf{s}) 
         =\frac{1}{\rm iDCG}\sum_{d=1}^K\frac{2^{\psi(d,\mathbf{s},\hat{\mathbf{s}}_{\theta})}-1}{\log_2(1+d)}.
    \end{aligned}
\end{equation}
The calculation of $\psi(d,\mathbf{s},\hat{\mathbf{s}}{\theta})$ involves sorting operations, making direct optimization of NDCG through gradient descent infeasible due to the non-differentiable nature of sorting. The sorting process can be formalized using a permutation matrix $\hat{\mathbf{s}}_{\theta}$, where each column $d$ indicates the position of each element of $\hat{\mathbf{s}}_{\theta}$ in the sorted order. Specifically, if the $j$-th element of $\hat{\mathbf{s}}_{\theta}$ is to be placed in the $d\rm{-th}$ position, then the entry $(j, d)$ in $\mathbf{P}_{\rm{hard}}$ is set to 1, and all other entries in the $d$-th column are set to 0. Thus, $\psi(d,\mathbf{s},\hat{\mathbf{s}}_{\theta})$ can be expressed as $\psi(d,\mathbf{s},\hat{\mathbf{s}}_{\theta}) = [\mathbf{P}_{\rm{hard}}^{\top} \cdot \mathbf{s}]_d $. To address this issue, a differentiable permutation matrix $\mathbf{P}_{\rm soft}$ can be obtained through the differentiable sorting network based on $\hat{\mathbf{s}}_{\theta}$, as defined in Eq. \ref{eq:P_soft}. With $\mathbf{P}_{\rm soft}$, the relaxed, differentiable score $\psi '(d,\mathbf{s},\hat{\mathbf{s}}_{\theta})$ at ranking position $d$ can be computed as:
\begin{equation*}
    \psi '(d,\mathbf{s},\hat{\mathbf{s}}_{\theta}) = [\mathbf{P}^{ \top }_{\rm soft} \cdot \mathbf{s}]_{d}.
\end{equation*}
Therefore, by substituting $\psi(d,\mathbf{s},\hat{\mathbf{s}}_{\theta})$ with $\psi '(d,\mathbf{s},\hat{\mathbf{s}}_{\theta})$, we can define our diffNDCG as follows: 
\begin{equation}
    \label{eq:loss_diffndcg}
         \mathrm{diffNDCG}(\hat{\mathbf{s}}_{\theta},\mathbf{s}) =\frac{1}{\rm iDCG}\sum_{d=1}^K\frac{2^{\psi '(d,\mathbf{s},\hat{\mathbf{s}}_{\theta})}-1}{\log_2(1+d)}.
\end{equation}

Finally, we consider the following optimization problem:
\begin{equation}
\label{final_objective}
     \min_{\theta} \mathcal{L}_{\rm diff} = \frac1{|D|}\sum_{(x,\mathbf{y},\mathbf{s})\in D} \ell_{\rm diff}(\mathbf{M}_{\rm ad}(x,\mathbf{y},\pi_{\theta}), \mathbf{s}),
\end{equation}
where
$
\ell_{\rm diff}(\cdot,\cdot) = -\mathrm{diffNDCG}(\cdot,\cdot).
$ The pseudo code of DRPO is presented in Algorithm~\ref{algorithm:drpo}.

\begin{table*}[t]
\begin{center}
\caption{Comparisons between our method and baselines. We report GPT-4 Win Rate (vs Chosen) and Reward Model Win Rate (vs Chosen and SFT). All Learning To Rank methods use PRR score.}
\label{tab:hh-qwen} 

\resizebox{\textwidth}{!}{%
    \begin{tabular}{l ccc | ccc }
    \toprule
    \multicolumn {1}{l}{ \bf {Base Model}}  
    &\multicolumn{3}{c|}{\bf {Qwen1.5-0.5B}}
    &\multicolumn{3}{c}{\bf {Qwen1.5-1.8B}}\\
    \bf Method  &\footnotesize{GPT-4 Win Rate}$\uparrow$ 
                &\footnotesize{RM Win Rate (vs Chosen)}$\uparrow$ 
                &\footnotesize{RM Win Rate (vs SFT)}$\uparrow$   
                &\footnotesize{GPT-4 Win Rate}$\uparrow$
                &\footnotesize{RM Win Rate (vs Chosen)}$\uparrow$ 
                &\footnotesize{RM Win Rate (vs SFT)}$\uparrow$
                \\
    \midrule    
    \multicolumn{7}{c}{ \small \bf {Preference Alignment Methods} }\\
    \midrule    
    SFT                     & 20.27\%($\pm$ 5.73) & 29.49\%($\pm$ 2.17)  & - 
                            & 37.43\%($\pm$ 4.48) & 30.66\%($\pm$ 1.86)   & -                   \\
    DPO                     & 28.26\%($\pm$ 3.40) & 34.96\%($\pm$ 1.69)  & 59.18\%($\pm$ 4.64) 
                            & 47.60\%($\pm$ 3.30) & 57.62\%($\pm$ 2.02)  & 76.06\%($\pm$ 3.67)   \\
    DPO\textsubscript{BT}   & 33.90\%($\pm$ 1.04) & 45.70\%($\pm$ 2.73)  & 69.72\%($\pm$ 4.53)
                            & 56.15\%($\pm$ 1.64) & 66.99\%($\pm$ 3.19)  & 81.84\%($\pm$ 1.01)   \\
    DPO\textsubscript{PL}   & 35.65\%($\pm$ 6.59) & 46.88\%($\pm$ 2.41)  & 71.29\%($\pm$ 2.09) 
                            & 55.09\%($\pm$ 4.69) & 63.67\%($\pm$ 1.41)  & 78.12\%($\pm$ 3.22)   \\
    PRO                     & 28.37\%($\pm$ 3.34) & 34.57\%($\pm$ 3.04)  & 56.64\%($\pm$ 3.00) 
                            & 37.59\%($\pm$ 4.91) & 45.12\%($\pm$ 2.16)  & 62.89\%($\pm$ 3.89)   \\
    LiPO                    & 35.59\%($\pm$ 4.28) & \underline{53.71\%}($\pm$ 2.49)  & \underline{79.10\%}($\pm$ 3.04) 
                            & \underline{62.95\%}($\pm$ 2.58) & \underline{73.63\%}($\pm$ 2.84)  & \underline{86.33\%}($\pm$ 2.50)   \\
    \midrule   
        \multicolumn{7}{c}{ \small \bf {Learning To Rank Methods} }\\
    \midrule    
    
    ListNet                 & 26.81\%($\pm$ 3.85) & 36.13\%($\pm$ 1.50)  & 60.94\%($\pm$ 2.34) 
                            & 42.74\%($\pm$ 3.31) & 49.80\%($\pm$ 1.69)  & 65.82\%($\pm$ 1.50)    \\
    PiRank                  & 26.08\%($\pm$ 2.61) & 38.87\%($\pm$ 2.73)  & 62.50\%($\pm$ 3.07) 
                            & 50.71\%($\pm$ 2.39) & 56.64\%($\pm$ 0.87)  & 69.14\%($\pm$ 2.10)    \\
    Neural Sort             & 26.40\%($\pm$ 4.30) & 35.35\%($\pm$ 2.49)  & 60.55\%($\pm$ 2.84) 
                            & 39.94\%($\pm$ 1.56) & 42.77\%($\pm$ 5.16)  & 58.00\%($\pm$ 5.16)    \\
    Fast Soft Sort          & \underline{37.58\%}($\pm$ 5.72) 
                                                  & 51.95\%($\pm$ 2.10)  & 74.99\%($\pm$ 2.34) 
                            & 61.90\%($\pm$ 7.68) & 71.87\%($\pm$ 3.08)  & 85.93\%($\pm$ 2.53)    \\
    Diff Sorting    & 28.97\%($\pm$ 1.46) & 44.53\%($\pm$ 1.65)  & 66.41\%($\pm$ 3.78)               
                            & 49.39\%($\pm$ 2.55) & 64.06\%($\pm$ 1.75)  & 78.56\%($\pm$ 2.31)     \\
    \midrule 
    \rowcolor{LightCyan} DRPO 
                            &\textbf{42.80\%($\pm$5.01)}
                            &\textbf{58.40\%($\pm$2.94)}
                            &\textbf{79.88\%($\pm$3.92)}
                            &\textbf{69.08\%($\pm$3.33)}
                            &\textbf{82.61\%($\pm$0.64)} 
                            &\textbf{89.06\%($\pm$2.40)} \\
    \bottomrule
    \end{tabular}
}
\vspace{-2em}
\end{center}
\end{table*}

\section{Experiments}

\label{sec:exp}

\subsection{Experimental Settings}

\textbf{Datasets.} \textbf{Anthropic's Helpful and Harmless} (\textbf{HH}) \cite{bai2022training} contains 161k/8.5k training/test samples. Each sample consists of a prompt and a pair of responses (chosen and reject), where ``chosen" represents the preferred response and ``reject" represents the less preferred response. We also generate additional responses for each prompt and rate each response using a reward model DeBERTa \footnotemark
\footnotetext{https://huggingface.co/OpenAssistant/reward-model-deberta-v3-large-v2}, resulting in ranking preference data of a list size $K=8$. For more details, please refer to Appendix. \textbf{UltraFeedback} \cite{cui2023ultrafeedback} contains 64k prompts. Each prompt corresponds to four responses, and every response has a score annotated by AI (e.g., GPT-4 and Gemini \cite{2023GPT4VisionSC,geminiteam2024gemini}). \textbf{VLFeedback} \cite{2023vlfeedback} consists of 80k multi-modal samples from various sources.

\textbf{Models.}
Our experiments are mainly based on Qwen1.5 model \cite{bai2023qwen} with a range of parameters from 0.5B to 1.8B and Mistral model \cite{jiang2023mistral} with 7B parameter size. In addition, we also train Qwen-VL-Chat \cite{bai2023qwenvl}, a large-scale vision-language model, to evaluate the performance of our method on multi-modal preference alignment task \cite{sun2023aligning}. 

\textbf{Baseline Methods.} To validate the effectiveness of our method, we conduct comparison experiments with representative baselines. In our experiments, we mainly compare our method with SFT, DPO \cite{rafailov2023direct}, PRO \cite{song2023preference}, LiPO \cite{liu2024lipo}, DPO\textsubscript{BT} and DPO\textsubscript{PL} \cite{rafailov2023direct}. DPO\textsubscript{BT} adapts DPO to ranking preferences by decomposing ranked lists into pairwise comparisons, as proposed by \cite{liu2024lipo}. Furthermore, DPO\textsubscript{PL} is proposed based on ranking preference data by \cite{rafailov2023direct}, which leverages the Plackett-Luce preference model \cite{af5079a1-8ca5-3727-a405-0a82390327b7}, a generalization of the Bradley-Terry model \cite{19ff28b9-64f9-3656-ba40-08326a05748e} that accommodates full rankings rather than just pairwise comparisons.

Furthermore, we implemented several differentiable sorting algorithms such as Fast Soft Sort \cite{DBLP:conf/icml/BlondelTBD20}, Neural Sort \cite{DBLP:conf/iclr/GroverWZE19} and learning-to-rank methods, including ListNet \cite{DBLP:conf/icml/XiaLWZL08}, PiRank \cite{DBLP:conf/nips/SwezeyGCE21} for list preference alignment. 
Unless otherwise specified, the learning-to-rank methods calculate scores using the Policy Reference Ratio Score (PRR).



\begin{table}[h]
    \begin{center}
    \caption{Performance comparison across different methods on Qwen models. RM and RM* represent DeBERTa and QRM-Llama3.1-8B-v2 reward models respectively. DRPO achieves superior performance across all evaluation metrics.}
    \resizebox{0.99\linewidth}{!}{%
        \begin{tabular}{lccccc}
        \toprule
        \textbf{Method} & \textbf{RM Score} & \textbf{RM* Score} & \textbf{MMLU} & \textbf{GSM8K} & \textbf{TruthfulQA}\\
        \midrule[\heavyrulewidth]
        \multicolumn{6}{c}{\bf {Qwen1.5-0.5B}} \\
        \midrule
        SFT & -1.7519 & 0.5204 & 39.25 & 14.78 & 38.64\\
        PRO & -1.4473 & 0.5449 & 36.19 & 14.85 & 38.26\\
        Fast Soft Sort & -0.7946 & 0.6109 & 38.82 & 18.45 & 37.77\\     
        LiPO & -0.6350 & 0.6244 & 40.26 & 17.36 & 38.87\\
        \midrule
        \rowcolor{LightCyan} DRPO
            & \textbf{-0.5195} & \textbf{0.6263} & \textbf{41.34} & \textbf{18.65} & \textbf{39.51}\\
        \midrule[\heavyrulewidth]
        \multicolumn{6}{c}{Qwen1.5-1.8B} \\
        \midrule
        SFT & -1.4651 & 0.5346 & 46.02 & 31.61 & 40.48\\
        PRO & -1.1162 & 0.5834 & 43.06 & 30.78 & 40.55\\
        Fast Soft Sort & -0.1291 & 0.6779 & 44.87 & 33.51 & 41.64\\
        LiPO & -0.0346 & 0.6891 & 46.57 & 33.88 & 42.41\\
        \midrule
        \rowcolor{LightCyan} DRPO & \textbf{0.5372} & \textbf{0.7082} & \textbf{47.32} & \textbf{35.14} & \textbf{44.67}\\
        \bottomrule
        \end{tabular}
    }
    \label{tab:gene_bench}
\end{center}
\end{table}

\begin{table}[t]
    \centering
    \caption{Performance comparison on LLama3-8B across different methods. We evaluate win rates (vs Chosen) and absolute scores using two reward models: RM (DeBERTa) and RM* (QRM-LLaMA3.1-8B-v2).}
    \label{tab:llama3_results}
    \resizebox{0.99\linewidth}{!}{
        \begin{tabular}{lcccc}
            \toprule
            \textbf{Method} & \textbf{RM Win Rate} & \textbf{RM Score} & \textbf{RM* Win Rate} & \textbf{RM* Score} \\
            \midrule[\heavyrulewidth]
            \multicolumn{5}{c}{LLama3-8B} \\
            \midrule
            SFT & 44.33\% & -1.1930 & 53.12\% & 0.5921 \\
            LIRE & 78.32\% & -0.0737 & 79.28\% & 0.6996 \\
            LiPO & 82.14\% & 0.3142 & 83.72\% & 0.7121 \\
            \midrule
            \rowcolor{LightCyan} DRPO & \textbf{85.35\%} & \textbf{0.5476} & \textbf{86.32\%} & \textbf{0.7333} \\ 
            \bottomrule
        \end{tabular}
    }
\end{table}

\begin{table}[t]
    \centering
    \caption{Comparison of PRR replacement with ARP Score across ranking methods. While ARP improves the baseline methods, DRPO still maintains better performance.}
    \label{tab:hh-qwen-arp}
    \resizebox{0.98\columnwidth}{!}{%

        \begin{tabular}{l c c}
            \toprule
            \bf \footnotesize{Method}  
                & \bf \footnotesize{ GPT-4 Win Rate}$\uparrow$ 
                & \bf \footnotesize{ RM Win Rate (vs Chosen)}$\uparrow$\\
            \midrule
            \midrule[\heavyrulewidth]
            \multicolumn{3}{c}{{Qwen1.5-0.5B}}                          \\
            \midrule
            \small{DPO\textsubscript{BT}(K=2)}        & 28.26\%   & 34.96\%     \\
            \footnotesize DPO\textsubscript{BT}(K=2) + ARP   & 31.72\%(+3.46)    & 43.55\%(+8.59)     \\
            \small PiRank              & 26.08\%                & 38.87\%                          \\
            \small PiRank + ARP        & 42.35\%(+16.27)        & 57.03\%(+18.16)           \\
            \small Fast Soft Sort      & 37.58\%                & 51.95\%                             \\
            \small Fast Soft Sort + ARP& 40.69\%(+3.11)         & 54.49\%(+2.54)            \\
            \midrule
            \rowcolor{LightCyan} DRPO
                                        & \textbf{42.80\%}               & \textbf{58.40\%}                  \\
            \midrule
            \midrule[\heavyrulewidth]
            \multicolumn{3}{c}{{Qwen1.5-1.8B}}  \\
            \midrule
            \small DPO\textsubscript{BT}(K=2)         & 47.60\%    & 57.62\%     \\
            \small DPO\textsubscript{BT}(K=2) + ARP   & 50.37\%(+2.77)          & 61.37\%(+3.75)             \\
            
            \small PiRank              & 50.71\%    & 56.64\%    \\
            \small PiRank + ARP        & 66.56\%(+15.85)         & 75.69\%(+18.95)     \\
            \small Fast Soft Sort      & 61.90\%    & 71.87\%      \\
            \small Fast Soft Sort + ARP& 66.14\%(+4.24)    & 77.92\%(+6.05)    \\
            \midrule
            \rowcolor{LightCyan} DRPO
                                & \textbf{69.08\% }   & \textbf{82.61\%}    \\
            \bottomrule
            
        \end{tabular}
    }
    \vspace{-1.25em}
\end{table}

\textbf{Implementation details.}

For experiments on the \textbf{HH} dataset, we use a learning rate of $5 \times 10^{-7}$. For the UltraFeedback and VLFeedback datasets, we employ a learning rate of $5 \times 10^{-6}$. Across all experiments, we adopt a global batch size of 32, a cosine learning rate schedule with a warmup ratio of 0.1, and the RMSprop optimizer~\cite{ruder2016overview}. All models are trained for one epoch on the entire training split. We set the maximum prompt length to 512 and the maximum context length to 1024. For LLaMA-3-8B~\cite{touvron2023llama} and Mistral-7B\footnote{\url{https://huggingface.co/HuggingFaceH4/mistral-7b-sft-beta}}, we utilize QLoRA~\cite{DBLP:conf/nips/DettmersPHZ23} to reduce memory consumption and facilitate efficient training.
We configure our diffNDCG loss function with the following default hyperparameters: ranking constant $\tau=0.2$, coefficient $\beta = 1$, and update rate $\theta = 0.9999$ for the Adaptive Rank Policy Score (Eq.\ref{eq:gamma_y_2}), and steepness $\alpha = 1$(Eq.\ref{eq:stepness}) for the differentiable sorting network. For DPO-style methods including DPO, DPO\textsubscript{BT}, and LiPO, we follow the same hyperparameter settings as in DPO~\cite{rafailov2023direct}, with $\beta = 0.1$.

\textbf{Evaluation.}
Our experiments use various metrics to evaluate the performance of different methods.

\begin{itemize}
    \item \textbf{RM Win Rate:} we use a trained reward model, such as DeBERTa, to evaluate the win rate of the generated response compared to either the preferred response within the dataset or the SFT target, where the SFT target is the response produced by the SFT model.
    \item \textbf{GPT-4 based Win Rate:} we use GPT-4 to compare which of generated responses is more preferred, and measure performance by win rate of responses.
    \item \textbf{Open Benchmarks:}
        We evaluate different methods on the UltraFeedback dataset using AlpacaEval2.0~\cite{alpaca_eval} and MT-Bench~\cite{DBLP:conf/nips/ZhengC00WZL0LXZ23} benchmarks. For the VLFeedback dataset, we employ MME~\cite{fu2023mme}, MMBench~\cite{liu2023mmbench}, and MMVet~\cite{Yu2023MMVetEL} benchmarks. Additionally, we assess performance on general benchmarks including MMLU~\cite{hendryckstest2021}, GSM8K~\cite{cobbe2021gsm8k}, and TruthfulQA~\cite{lin2022truthfulqameasuringmodelsmimic}.

\end{itemize}

\begin{table}[t]
    \caption{Ablation experiments of individual components.}

    \centering
    \label{tab:ablation}
    \resizebox{0.95\columnwidth}{!}{%
        \begin{tabular}{l c c c}
            \toprule
            \bf \small{Method}  
            & \bf \small{GPT-4 Win Rate$\uparrow$ }
            & \bf \small{RM Win Rate (vs Chosen)$\uparrow$} \\
            \midrule
            DRPO-w/o ada \& diff         & 28.97\%($\pm$1.46)  & 44.53\%($\pm$1.65) \\
            DRPO-w/o diff                & 33.06\%($\pm$5.33)  & 47.66\%($\pm$3.49) \\
            DRPO-w/o ada                 & 38.30\%($\pm$3.35)  & 53.91\%($\pm$1.46) \\
            \midrule
            \rowcolor{LightCyan} DRPO
                & \textbf{42.80\%($\pm$5.01)} 
                & \textbf{58.40\%($\pm$2.94)}\\
            \bottomrule
        \end{tabular}%
    }
    \vspace{-1.0em}
\end{table}

\subsection{Experiments Results on Anthropic's Helpful and Harmless Dataset}  

\textbf{Main Results.} Experiments conducted on the HH dataset are presented in Table \ref{tab:hh-qwen}, demonstrating the effectiveness of our method. Our method, DRPO, outperforms baseline methods including SFT, DPO\textsubscript{BT}, DPO\textsubscript{PL}, PRO, LiPO, and other LTR methods across different model scales. The GPT-4 Win Rate has an improvement of $5.22\% {\sim} 6.13\%$, and Reward Model Win Rate has an improvement of $4.69\% {\sim} 8.98\%$ and $0.78\% {\sim} 2.73\%$. Several key observations emerge from these results. First, even without additional modifications, our Sorting Network outperforms most conventional sorting methods, such as Neural Sort. Second, directly extending existing alignment methods (e.g., DPO) to ranking list scenarios yields performance improvements. Third, while LiPO employs a metric weighting scheme to account for the relative importance of responses at different positions and achieves superior performance compared to DPO\textsubscript{BT}, DRPO leverages diffNDCG to precisely quantify response contributions at each ranking position, resulting in significantly enhanced performance over LiPO. Additionally, Appendix presents a comparison of reward distributions for model-generated responses, revealing improvements in response rewards when using our method.

\begin{table}[t]
    \centering
    \caption{Effect of Discount Factor Selection on diffNDCG.}
    \label{tab:hh-qwen-discount}
    \resizebox{0.95\columnwidth}{!}{%
        \begin{tabular}{c c c c}
            \toprule
            \bf \small{Discounts}  & \bf \small{GPT-4 Win Rate}$\uparrow$ & \bf \small{ RM Win Rate (vs SFT)}$\uparrow$ & \small{Average}$\uparrow$ \\
            \midrule
            $1/{\sqrt{r}}$  & 40.53\%($\pm$4.36)       & 79.29\%($\pm$2.89) & 59.91\%\\
            $1/r$           & \textbf{43.37}\%($\pm$5.14)       & 78.32\%($\pm$2.94) & 60.84\%\\
            $1/{r^2}$       & 40.37\%($\pm$4.17)       & 79.30\%($\pm$3.38) & 59.83\%\\
            \midrule
            \rowcolor{LightCyan} $1/{\mathrm{log}(1+r)}$ 
                            & 42.80\%($\pm$5.01)       & \textbf{79.88}\%($\pm$3.72)  & \textbf{61.34}\% \\
            \bottomrule
        \end{tabular}
    }
\end{table}
\begin{table}[t]
    \caption{Time complexity and memory usage comparisons.}
    \label{tab:hh-qwen-cost}
    \vspace{-0.1em}
    \centering
    \resizebox{0.95\columnwidth}{!}{%
        \begin{tabular}{l cc | cc }
        \toprule
        \bf \small{Method}  & \bf \small{Time Complexity}
                    & \bf \small{Run Time}
                    & \bf \small{Memory}  \\
        \midrule
        PRO            & $\mathcal{O}(L^2)$  & 0.2806s  & 24.57GB \\
        LiPO           & $\mathcal{O}(L^2)$  & 0.3269s  & 24.08GB  \\
        Neural Sort    & $\mathcal{O}(L^2)$  & 0.2585s  & 24.01GB   \\
        Pirank         & $\mathcal{O}(L^2)$  & 0.2635s  & 25.80GB   \\
        Fast Soft Sort & $\mathcal{O}(L \log (L))$ & 0.2618s  & 23.21GB   \\
        ListNet        & $\mathcal{O}(L^2)$  & 0.2652s  & 21.10GB   \\
        \midrule
        \rowcolor{LightCyan} DRPO (odd-even) &  $\mathcal{O}(L^2)$
                                        & 0.2641s
                                        & 23.39GB \\
        \rowcolor{LightCyan} DRPO (bitonic) &  $\mathcal{O}(L\log^2(L))$
                                        & 0.2560s
                                        & 23.43GB \\
        \bottomrule
        \end{tabular}
    }
    \vspace{-1.0em}
\end{table}

\textbf{Validation on Multiple Reward Models and Open Benchmarks.} To comprehensively validate our method, we employ multiple reward models and evaluation metrics. Specifically, we use QRM-Llama3.1-8B-v2~\cite{dorka2024quantile}, a top-performing model in RewardBench~\cite{lambert2025rewardbench}, alongside the DeBERTa-based reward model. We evaluate on three established benchmarks: MMLU~\cite{hendryckstest2021}, GSM8K~\cite{cobbe2021gsm8k}, and TruthfulQA~\cite{lin2022truthfulqameasuringmodelsmimic}. 

Table~\ref{tab:gene_bench} presents comprehensive comparisons with competitive baselines. "RM" and "RM*" denote evaluations using DeBERTa and QRM-Llama3.1-8B-v2 respectively, while "Score" represents the absolute reward score. DRPO consistently outperforms all baselines across all metrics, with consistent preferences demonstrated by both reward models, validating the effectiveness of our approach.

\textbf{Comparison on larger LLMs.} We validate our method on a larger-scale LLM (LLaMA3-8B~\cite{touvron2023llama}). We further compare our approach with an additional ranking-based method LIRE~\cite{acl_lire} and a strong baseline from previous experiments, LiPO~\cite{liu2024lipo}. Results are presented in Table~\ref{tab:llama3_results}. While LIRE and LiPO both improve over the SFT baseline, our method consistently outperforms all competing approaches across different model scales. This demonstrates the effectiveness and scalability of our approach regardless of model size.


\subsection{Ablation Studies}

\textbf{Ablation Studies on Adaptive Rank Policy Score.} To assess the efficacy of our proposed Adaptive Rank Policy Score, we conducted a comparative analysis by integrating it into multiple baseline models. The results are presented in Table \ref{tab:hh-qwen-arp}. Our experimental results demonstrate that the Adaptive Rank Policy Score significantly enhances baseline model performance across various sizes, showcasing the advantages of applying ARP over the PRR score.

Furthermore, we conduct an ablation study on our method by progressively removing Adaptive Rank Policy Score(ARP) and diffNDCG, denoted as 'w/o ARP' and 'w/o diffNDCG' respectively. As shown in Table \ref{tab:ablation}, we observe that directly using the cross-entropy loss between the ground truth permutation matrix $\mathbf{P}_{\rm ground}$ and the predicted permutation matrix $\mathbf{P}_{\rm soft}$, without ARP and diffNDCG, leads to performance degradation. This fails to account for the varying importance of responses at different ranking positions. Additionally, when the ARP is replaced by PRR in Eq \eqref{m1_dpo}, performance decreases by $4.09\%-4.5\%$ and 3.13\%-4.49\% for GPT-4 Win Rate and RM Win Rate, respectively.

 \textbf{Ablation Studies on DiffNDCG Discounts Factors.}

To further analyze the impact of discount factors on diffNDCG, we conducted ablation studies by comparing our default discount factors with alternative variants. Specifically, we set up three  variants for comparison: inverse square root ($1/\sqrt{r}$), inverse linear ($1/r$), and inverse quadratic ($1/r^2$), where $r$ denotes the ranking position. Our default formulation employs the logarithmic discount factor $1/\log(1+r)$. The experimental results are presented in Table~\ref{tab:hh-qwen-discount}. 

We observed that various discount factors are indeed effective for diffNDCG. Among the examined variants, our default logarithmic discount ($1/\log(1+r)$) emerged as the most well-balanced choice, offering an effective compromise between emphasizing top-ranked responses and appropriately penalizing their misplacements. We hypothesize that this balance stems from the characteristics of the logarithmic discount function: excessively steep discount factors may cause models to entirely disregard lower-ranked responses, while overly gradual discount factors may fail to sufficiently penalize ranking position errors.

 
\begin{table}[t]
\begin{center}
\caption{The ablation study on sorting networks in DRPO suggests that odd-even sorting networks tend to demonstrate favorable performance compared to alternative architectures.}
\label{tab:hyper_sorting_network} 
\resizebox{0.99\linewidth}{!}{%
    \begin{tabular}{>{\centering}m{2cm}>{\centering}m{1.75cm}>{\centering}m{1.75cm}>{\centering\arraybackslash}m{0.75cm}}
        \toprule
        \textbf{Method} & \textbf{RM Winrate (vs Chosen)} & \textbf{RM Winrate (vs SFT)} & \textbf{Average}\\
       
    \midrule
    DRPO(bitonic)        & 59.17\%($\pm2.78$) & 75.58\%($\pm2.58$) & 67.37\% \\
    DRPO(odd-even)       & 58.98\%($\pm4.95$) & 78.90\%($\pm3.70$) & 68.94\% \\
    \bottomrule
    \end{tabular}
}
\vspace{-1.25em}
\end{center}
\end{table}

\begin{figure}[t]
    \centering
    \vspace{-1em}
    \includegraphics[width=0.9\columnwidth]{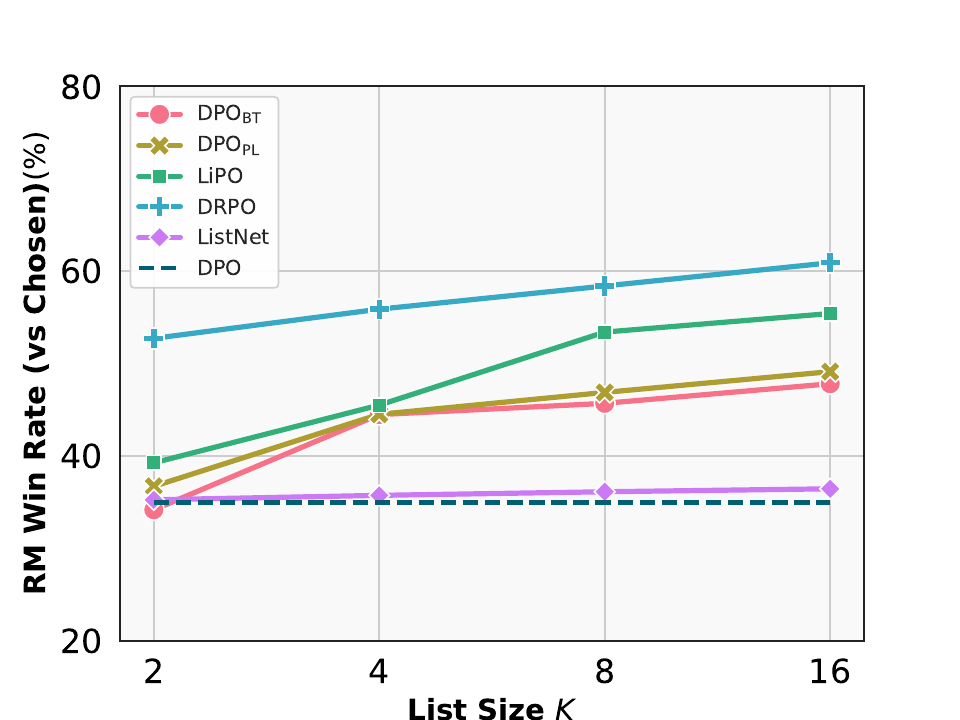}
    \caption{Method Comparison across different list lengths.}
    \label{fig:ablation_k}
    \vspace{-1.0em}
\end{figure}

\textbf{Ablation Studies on List Size.} We also analyze the performance for different list sizes $K$ of ranking preference data. We compared various list-wise methods and included DPO as a comparison baseline on the HH dataset, using Qwen1.5-0.5B as the base model. As Figure~\ref{fig:ablation_k} shows, our method consistently outperforms all baselines across different list sizes $K$ ranging from 2 to 16. Notably, all list-wise methods surpass DPO even at $K=2$, demonstrating the inherent advantage of list-wise optimization over pair-wise approaches.

\textbf{Time Complexity Analysis And Computational Efficiency.} We conducted efficiency experiments comparing different sorting algorithms, with results presented in Table \ref{tab:hh-qwen-cost}. The running times represent average loss computation time on 256 samples using the Qwen1.5-0.5B model. Our differentiable odd-even sorting network and diffNDCG computation both have time complexity $\mathcal{O}(L^2)$ for ranking lists of length $L$ (where each swap is $\mathcal{O}(1)$), comparable to Neural Sort, PiRank, and LiPO. Despite the quadratic complexity, our actual running times are comparable to Fast Soft Sort's $\mathcal{O}(L\log(L))$ implementation. While Fast Soft Sort has lower theoretical complexity, the small ranking list sizes in practice reduce the complexity gap's impact, making our method practically efficient.
\label{sec:time_complexity_analysis}

\textbf{Ablation Study on Different Sorting Networks.} Sorting networks constitute a fundamental architectural component of our approach. We investigate two prevalent sorting network variants: the odd-even sorting network, which performs pairwise comparisons and swaps at alternating odd and even indices, and the bitonic sorting network, which recursively constructs and merges bitonic sequences \cite{10.5555/533017}. As shown in Table \ref{tab:hyper_sorting_network}, the odd-even sorting network demonstrates a marginal performance advantage over the bitonic variant. Furthermore, complexity analysis (Table \ref{tab:hh-qwen-cost}) reveals comparable computational efficiency between these two architectures. Based on these empirical evaluations, we adopt the odd-even sorting network for our final DRPO implementation.

\begin{table}[t]
\begin{center}
\caption{Ablation Study on Hyperparameter Effects in DRPO}
\label{tab:hyper_eq5} 
\resizebox{0.99\linewidth}{!}{%
    \begin{tabular}{l c  c }
    \toprule
    \textbf{Method} & \textbf{RM Winrate(vs Chosen)} & \textbf{RM Winrate(vs SFT)} \\
    \midrule
    \multicolumn{3}{c}{\bf { Ranking Constant $\tau$}}\\
    \midrule
    DRPO($\tau=0.1$)        & 60.50\%($\pm2.62$) & 77.53\%($\pm3.42$) \\
    DRPO($\tau=0.2$)        & 58.98\%($\pm4.95$) & 78.90\%($\pm3.70$) \\
    DRPO($\tau=0.5$)        & 59.76\%($\pm2.73$) & 77.92\%($\pm2.78$) \\
    DRPO($\tau=1$)          & 52.34\%($\pm3.70$) & 72.07\%($\pm3.59$) \\

    \midrule
    \multicolumn{3}{c}{\bf { Coefficient $\beta$ }}\\
    \midrule
    DRPO($\beta=0.5$)        & 57.42\%($\pm1.40$) & 80.27\%($\pm1.50$) \\
    DRPO($\beta=1$)       & 58.98\%($\pm4.95$) & 78.90\%($\pm3.70$) \\
    DRPO($\beta=2$)        & 59.37\%($\pm4.17$) & 80.07\%($\pm1.70$) \\
    
    \midrule
    \multicolumn{3}{c}{\bf { Update Rate $\theta$}}\\
    \midrule
    DRPO($\theta=0.99$)        & 62.50\%($\pm2.92$) & 78.12\%($\pm2.59$) \\
    DRPO($\theta=0.999$)        & 59.96\%($\pm2.36$) & 78.71\%($\pm2.61$) \\
    DRPO($\theta=0.9999$)        & 58.98\%($\pm4.95$) & 78.90\%($\pm3.70$) \\
    \midrule

    \multicolumn{3}{c}{\bf { Steepness $\alpha$ }}\\
    \midrule
    DRPO($\alpha=0.1$)        & 58.39\%($\pm1.77$) & 76.36\%($\pm2.94$) \\
    DRPO($\alpha=1$)       & 58.98\%($\pm4.95$) & 78.90\%($\pm3.70$) \\
    DRPO($\alpha=10$)        & 59.76\%($\pm3.68$) & 79.10\%($\pm4.95$) \\
    DRPO($\alpha=50$)        & 58.59\%($\pm1.46$) & 78.71\%($\pm0.64$) \\
    \bottomrule
    \end{tabular}
}
\vspace{-1.25em}
\end{center}
\end{table}


\textbf{Ablation Study on Hyperparameters.} We conduct comprehensive ablation studies on the hyperparameters of two critical components: the Adaptive Rank Policy Score and differentiable sorting networks. Results in Table~\ref{tab:hyper_eq5} show consistent performance across different configurations, guiding our final hyperparameter selection.
\begin{itemize}
\item \textbf{Ranking Constant $\tau$:} This parameter controls the basic margin between adjacent responses in the ranking. Our experiments show that smaller values in the range $\tau \in [0.1, 0.5]$ maintain stable performance (58-60\% RM Win rate vs Chosen), while larger values ($\tau = 1$) lead to noticeable performance degradation. This suggests that moderate margins are sufficient for effective rank discrimination.

\item \textbf{Coefficient $\beta$:} This parameter determines the influence of the historical estimate $V_{q(y)}$ on the current score. The results demonstrate robust performance across the range $\beta \in [0.5, 2]$, with all values achieving comparable performance. This indicates flexibility in balancing current and historical information.

\item \textbf{Update Rate $\theta$:} This parameter controls the exponential moving average update rate for tracking historical changes. Values in the range $\theta \in [0.99, 0.9999]$ all yield strong performance (78-79\% RM Win rate vs SFT), though slightly higher update rates ($\theta = 0.99$) show marginal advantages on the RM Win rate (vs Chosen). This suggests a preference for moderately fast adaptation to recent training dynamics.

\item \textbf{Steepness $\alpha$:} This parameter controls the steepness of the s-shaped function in the differentiable sorting network's swapping operation. Our method exhibits robustness across an extremely wide range $\alpha \in [0.1, 50]$, with all configurations achieving nearly identical performance.
\end{itemize}

\begin{table}[t]
    \centering
    \caption{ Comparison of aligning Mistral-7B-Base on the UltraFeedback dataset: Our DRPO outperforms others. }
    \label{tab:mistal-ultrafeedback}
    \resizebox{0.99\columnwidth}{!}{%
        \begin{tabular}{c c c c }
        \toprule
            \bf \small{Method} 
            & \bf \small{MT-Bench $\uparrow$ }
            & \bf \small{AlpacaEval\textsubscript{2.0}(LC) $\uparrow$}
            & \bf \small{AlpacaEval\textsubscript{2.0}(WR) $\uparrow$}\\
        \midrule
        SFT     & 6.3  & 8.4\%     & 6.2\% \\
        DPO     & 7.3   & 15.1\%    & 12.5\%\\
        ORPO    & 7.3   & 14.7\%    & 12.2\% \\  
        R-DPO   & 7.4   & 17.4\%    & 12.8\%        \\
        LiPO    & \textbf{7.4}  & 25.3\%    & 18.9\% \\
        \midrule
        \rowcolor{LightCyan} DRPO
                & 7.3 & \textbf{26.5\%} &  \textbf{19.6\%} \\
        \bottomrule
        \end{tabular}
    }
    \vspace{-0.5em}
\end{table}

\begin{table}[t]
    \centering
    \caption{Comparisons between our method and baselines on multi-modal VLFeedback dataset.}
    \label{tab:qwen-vl-chat-vlfeedback}
        
    \resizebox{0.85\linewidth}{!}{%
    
        \begin{tabular}{p{1cm} c c c }
        \toprule
            \bf \small{Model}  
            & \bf \small{MME\textsuperscript{P} $\uparrow$ }
            & \bf \small{MM-Bench $\uparrow$}
            & \bf \small{MM-Vet $\uparrow$}\\
        \midrule
        DPO                     & 1496.7 & 52.83\% &45.2 \\
        DPO\textsubscript{BT}   & 1548.1 & 52.55\% &46.8 \\
        LiPO                    & 1559.3  & 54.55\% &47.2  \\
        \midrule
        \rowcolor{LightCyan} DRPO 
                                & \textbf{1581.1} & \textbf{56.19\%} &\textbf{48.6} \\
        \bottomrule
        \end{tabular}            
    }
    \vspace{-0.5em}
\end{table}

\subsection{Experimental Results on UltraFeedback and VLFeedback Dataset}

\textbf{Main Results on UltraFeedback Dataset and VLFeedback Dataset.} To validate scalability and performance, we train a Mistral model on the UltraFeedback dataset and evaluate it using open benchmarks. As shown in Table \ref{tab:mistal-ultrafeedback}, our method can scale up to larger model and outperform other methods. In MT-Bench, we achieve improvements of 1.2\% in AlpacaEval\textsubscript{2.0} length-controlled Win Rate (LC) and 0.7\% in AlpacaEval\textsubscript{2.0} raw Win Rate (WR). We aslo apply DRPO to fine-tune vision-language models. We use Qwen-VL-Chat as our base model and evaluate each method with multi-modal benchmarks. As Table \ref{tab:qwen-vl-chat-vlfeedback} shows, our method outperform other methods. Our method demonstrates improved performance across multiple benchmarks, achieving improvements of $21.8$ in MME perception tasks, $1.4$ in MM-Vet, and $1.64\%$ in MM-Bench.



\section{Limitations and Conclusion}

 Aligning LLMs with human preferences is crucial for enhancing interactions and the safety of LLMs. We propose a novel method  DRPO, which treats human preference alignment as a listwise ranking problem and aligns LLMs using ranking preference data. Specifically, we introduce an Adaptive Rank Policy Score for ranking computation and develop a diffNDCG loss function based on the NDCG metric. Our extensive experimental results demonstrate the effectiveness of the proposed method, paving the way for future research. However despite using a large reward model as a proxy for human evaluations, discrepancies from actual human judgments may impact model performance. Generally, more sophisticated reward models provide more accurate evaluations. Future work could explore more sophisticated reward models to better approximate human preferences.

\bibliographystyle{IEEEtran}
\bibliography{refs}




\end{document}